# Gate-Controlled Quantum Dots Based on Two-Dimensional Materials

Fang-Ming Jing, Zhuo-Zhi Zhang, Guo-Quan Qin, Gang Luo, Gang Cao, Hai-Ou Li, Xiang-Xiang Song[*],

and Guo-Ping Guo[*]

*Mr. F.-M. Jing, Dr. Z.-Z. Zhang, Mr. G.-Q. Qin, Dr. G. Luo, Prof. G. Cao, Prof. H.-O. Li, Dr. X.-X. Song, Prof. G.-P. Guo*
CAS Key Laboratory of Quantum Information
CAS Center for Excellence in Quantum Information and Quantum Physics
University of Science and Technology of China, Hefei, Anhui 230026, China
Email: songxx90@ustc.edu.cn; gpguo@ustc.edu.cn

*Prof. G.-P. Guo*
Origin Quantum Computing Company Limited, Hefei, Anhui 230088, China

F.-M. Jing and Z.-Z. Zhang contributed equally to this work.

## Abstract

Two-dimensional (2D) materials are a family of layered materials exhibiting rich exotic phenomena, such as valley-contrasting physics. Down to single-particle level, unraveling fundamental physics and potential applications including quantum information processing in these materials attracts significant research interests. To unlock these great potentials, gate-controlled quantum dot architectures have been applied in 2D materials and their heterostructures. Such systems provide the possibility of electrical confinement, control, and manipulation of single carriers in these materials. In this review, efforts in gate-controlled quantum dots in 2D materials are presented. Following basic introductions to valley degree of freedom and gate-controlled quantum dot systems, the up-to-date progress in etched and gate-defined quantum dots in 2D materials, especially in graphene and transition metal dichalcogenides, is provided. The challenges and opportunities for future developments in this field, from views of device design, fabrication scheme, and control technology, are discussed. The rapid progress in this field not only sheds light on the understanding of spin-valley physics, but also provides an ideal platform for investigating diverse condensed matter physics phenomena and realizing quantum computation in the 2D limit.

## Table of Contents

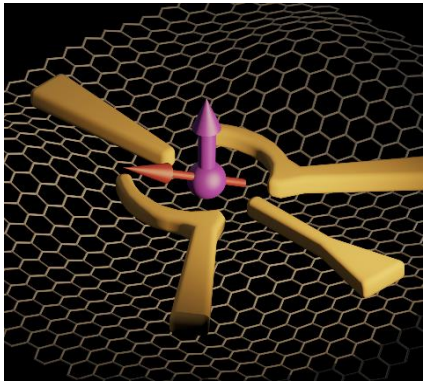

Gate-controlled quantum dots provide access to electrically manipulating individual charge carriers. This paper reviews recent advances of etched and gate-defined quantum dots based on two-dimensional materials. Such nanostructures offer promising platforms to study various quantum degrees of freedom at single-particle level, showing great potentials towards realization of solid-state quantum computation in two-dimensional materials.

# 1 INTRODUCTION

A quantum dot[1] is a small island that can be charged by electrons. Typically, a quantum dot connects to a source and a drain contact through tunneling barriers, which allows electrons to transfer through the quantum dot. Also, a quantum dot is usually coupled to several gate electrodes via capacitances. These gates can tune the energy levels of the quantum dot, as well as the tunneling barriers. When the size of the dot is comparable to the wavelength of its confined electrons (usually tens to hundreds of nanometers), the quantum dot exhibits a discrete energy spectrum, behaving as an artificial atom.

Through tuning the voltages applied to gate electrodes, electrons can be controlled to tunnel into the quantum dot one by one, due to Coulomb repulsion. Employing proper device architectures and gating strategies, quantum dots containing few electrons or even one electron can be achieved[2]. Thus gate-controlled quantum dot offers an electrically-tunable platform for studying electrons in solids down to single-particle level. With the powerful capability of electrical tuning, gate-controlled quantum dots can be used for simulating quantum phenomena[3-5] and studying mesoscopic physics[6-13]. Moreover, it also allows the possibilities for manipulating electron's charge and spin degrees of freedom[14-19], providing a promising way for realizing solid-state quantum computation[20]. Taking advantages of good coherency and convenience for integrations, tremendous progress has been made towards this goal[21-23], including realizing various types of qubits[24-34], integrating qubits with quantum bus[35-40], and demonstrating proof-of-principle quantum algorithms[41-43].

Meanwhile, since its discovery, graphene has exhibited extraordinary properties[44], not only attracting considerable research interests[45, 46], but also giving birth to studies of a relevant family of materials known as two-dimensional (2D) materials[47, 48]. 2D materials are a family of layered materials composed of atomically-thin layers vertically stacked together by van der Waals forces. Their dangling-bond-free interfaces allow realization of heterostructures based on materials of different properties in a 2D fashion using transfer techniques[49, 50]. Rich exotic phenomena have been demonstrated in 2D materials, such as massless Dirac fermions[51-53], viscous electron fluids[54-59], and superlattices induced by twisted stacking of adjacent flakes[60-67]. Among them, valley-contrasting physics are of particular interest[68-70]. The valley degree of freedom arises from degenerate energy extrema in the conduction or valence bands. Typically, different valleys are distantly separated in momentum space. As a consequence, the electron intervalley scattering is expected to be weak, leading to a long valley lifetime. Therefore, the valley degree of freedom offers a promising quantum degree of freedom that can be used to encode, process, and store information, serving as qubits for realizing quantum computation[71-73]. In order to fully unlock this potential, manipulating the valley degree of freedom at single-particle level is essential. Thus, it motivates realization of electrically confining carriers using nanostructures in 2D materials. Among them, gate-controlled quantum dots stand out. They can provide electrically tunable confinements of individual charge carriers, allowing electrostatic control on their charge, spin, and valley degrees of freedom.

We will start with brief introductions to valley degree of freedom in 2D materials and basic properties of gate-controlled quantum dots. Then the progress of quantum dots based on 2D materials will be reviewed, mainly focusing on two kinds of materials, graphene, and transition metal dichalcogenides (TMDs). These are the most well-studied material systems, due to their unique spin-valley coupling physics and potentials for quantum computation applications. Progress based on other 2D materials will also be discussed. Finally, we will conclude with an outlook on the challenges and opportunities of 2D material-based gate-controlled quantum dots. Previous reviews can be referenced. For example, reviews on valley-dependent properties and applications in 2D materials can be found in Refs.[68-70, 74]. Reviews on quantum dot systems, especially on the theory and applications in quantum computation, have been given in Refs.[75, 76]. For previous reviews on quantum dots based on 2D materials and their applications in quantum information processing, see Refs.[72, 73,

## 2 VALLEY DEGREE OF FREEDOM IN 2D MATERIALS

Taking molybdenum disulfide ($MoS_2$), one of the most representative TMDs materials, as an example[74], its monolayer has a honeycomb lattice, similar to graphene (Figure 1a). As shown in Figure 1b, both conduction and valance bands have extrema located at the corners of the first Brillouin zone. The six corners can be divided into two groups denoted by K and −K (also refers as K′) points respectively, forming two energetically degenerate but inequivalent valleys. K and −K valleys can be transformed to each other under time reversal operations. The two most important physical quantities for controlling valley degree of freedom are Berry curvature $\boldsymbol{\Omega}$ and orbital magnetic moment $\boldsymbol{\mu}_V$, which directly interact with applied electric and magnetic fields, respectively (Ref.[70]). Both of them are odd under time reversal (for example, $\boldsymbol{\Omega}(\boldsymbol{k}) = -\boldsymbol{\Omega}(-\boldsymbol{k})$, where $\boldsymbol{k}$ is the crystal momentum) and even under spatial inversion ($\boldsymbol{\Omega}(\boldsymbol{k}) = \boldsymbol{\Omega}(-\boldsymbol{k})$). As a result, in order to obtain valley-contrasting phenomena, inversion symmetry needs to be broken.

In monolayer $MoS_2$, due to the lack of inversion symmetry, carriers in K and −K valleys have opposite Berry curvatures $\boldsymbol{\Omega}$, which act like effective magnetic fields in the momentum space. As a consequence, valley-contrasting Hall current can be observed in the presence of an in-plane electric field (for bilayer $MoS_2$, a vertical electric field is needed to break the inversion symmetry), giving rise to the observation of the valley Hall effect[80-85]. In graphene systems, a similar valley Hall effect can be realized by breaking the inversion symmetry. This can be achieved by employing graphene/hexagonal boron nitride (hBN) superlattices for monolayer graphene[86, 87] and by applying an out-of-plane electric field for bilayer graphene[88-90] (Figure 1c). Breaking inversion symmetry also allows valley-contrasting orbital magnetic moments $\boldsymbol{\mu}_V$, which have opposite signs for the K and −K valleys. The orbital magnetic moments have the orientation of out-of-plane and interact with applied magnetic fields via a Zeeman-like interaction known as valley Zeeman effect. Therefore, it allows to lift the K and −K valley degeneracy by applying an out-of-plane magnetic field[91-95].

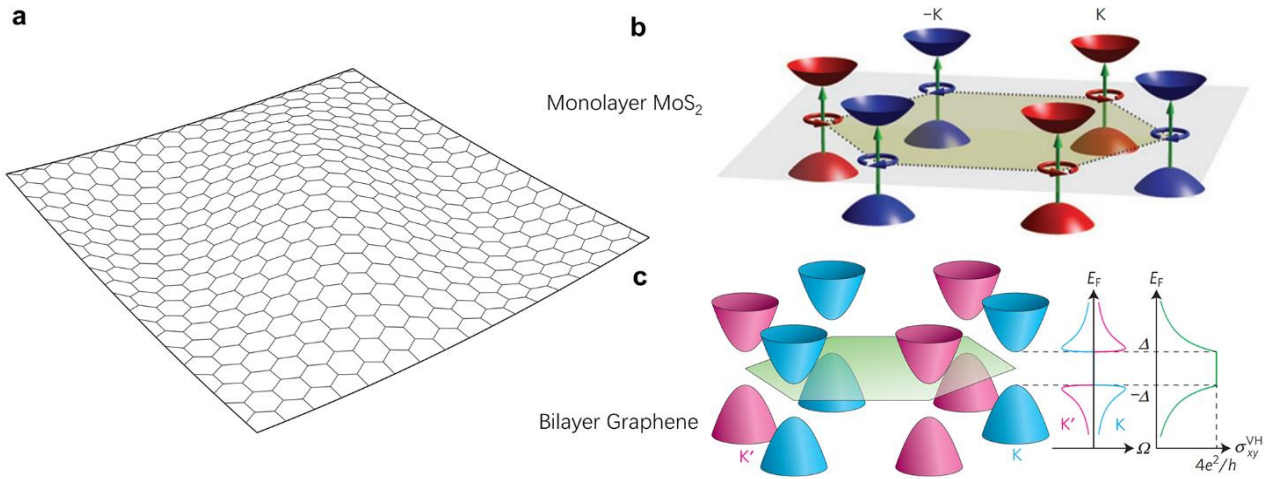

Figure 1. a) Schematic of two-dimensional hexagonal lattice. b) Band structure of monolayer $MoS_2$, exhibiting two energetically degenerate but inequivalent valleys. c) Band structure, Berry curvature, and valley Hall conductivity of bilayer graphene, when applying a perpendicular electric field to break the inversion symmetry. b) Reproduced with permission[68]. Copyright 2014, Springer Nature. c) Reproduced with permission[88]. Copyright 2015, Springer Nature.

Investigations on valley degree of freedom have attracted significant research interests in the past decade, leading to the rapid development of valleytronics[69]. Thanks to the valley-dependent optical selection rules[96-99], the valley degree of freedom is easy to access in TMDs. A variety of valley-dependent phenomena have been demonstrated and studied using optical approaches[70], showing great potentials for manipulating valley degree of freedom. However, electrical access to the valley degree of freedom is rather difficult. In addition to the (inverse) valley Hall effect mentioned above, an alternative way for electrical investigating valley degree of freedom is using Shubnikov-de Hass oscillations and quantum Hall effect, where properties of valley states can be examined under a perpendicular magnetic field[51, 52, 100-110]. As we will discuss below, gate-controlled quantum dots offer a highly tunable platform that allows electrically access and manipulate the valley degree of freedom in 2D materials.

## 3 GATE-CONTROLLED QUANTUM DOTS

A quantum dot system can be viewed as a network consisting of tunnel resistors and capacitors. Figure 2a shows schematics of a single quantum dot (upper panel) and a double quantum dot coupled in series (lower panel). Taking single dot as an example, the quantum dot (QD) is coupled to source (S) and drain (D) contacts via tunneling barriers, respectively. The tunneling barriers can be modeled by a tunnel resistor and a capacitor, as indicated in the inset. Meanwhile, the quantum dot is capacitively coupled ($C_G$) to a gate electrode, which allows applying gate voltage ($V_G$) for electrical tuning. Due to the small size of confinement, the energy spectrum becomes discrete.

We usually use the constant interaction model to describe the behaviors of quantum dots[1, 76]. In this model, two basic assumptions are conducted. 1) Coulomb interactions among electrons are characterized by a single and constant capacitance $C = C_S + C_D + C_G$, where $C_S$, $C_D$, $C_G$ are capacitances between the dot and the source, the drain, and the gate, respectively. 2) The single-particle energy level is independent of the number of electrons in the dot. Under these assumptions, we can derive the electrochemical potential $\mu(N)$. It describes the energy cost for adding the $N$th electron to the quantum dot which contains $N - 1$ electrons and is given by $\mu(N) = (N - N_0 - 1/2)E_C - E_C(C_S V_S + C_D V_D + C_G V_G)/|e| + E_N(B)$. Here, $N_0$ is the number of electrons responsible for the background charge. $E_C = e^2/C$ is the charging energy. $V_S$, $V_D$, $V_G$ are voltages applied to the source, the drain, and the gate, respectively. $E_N(B)$ is the single-particle energy level of the $N$th electron, which can be manipulated by the magnetic field $B$. Note that $\mu(N)$ linearly depends on the gate voltage and the dependence is same for all $N$. This results in the electrochemical potential "ladder", as shown in Figure 2b. The spacing of the "ladder" is given by $\mu(N + 1) - \mu(N) = E_C + \Delta E$, known as addition energy $E_{add}$. Here $E_{add}$ has two parts of contribution: one is from the charging energy $E_C$, the other is from $\Delta E = E_{N+1} - E_N$, which is the energy spacing between two discrete quantum levels. Usually $\Delta E$ is much smaller than $E_C$ if the dot contains a large number of electrons. When the dot enters in few-electron regime, the contribution of $\Delta E$ reveals.

From extracting $\Delta E$, we can get access to the quantum states and study their properties. This can be achieved through transport measurements. As shown in the left panel of Figure 2b, if a particular electrochemical potential $\mu(N)$ is tuned into the bias window ($\mu_S \geq \mu(N) \geq \mu_D$, where $\mu_S$ and $\mu_D$ are electrochemical potentials of the source and the drain reservoir, respectively), an electron can tunnel through the dot, developing a current to be measured (in particular, the single-electron tunneling event occurring when $\mu_S$, $\mu_D$, and $\mu(N)$ are aligned is referred as resonant tunneling). Otherwise, no current flows, which is known as Coulomb blockade (right panel of Figure 2b). Continuing tuning the electrochemical potential ladder by varying $V_G$, a series of current peaks can be observed, known as Coulomb oscillations (lower panel of Figure 2c). At each peak, an electron is loaded into the dot and then tunnels out, while between the peaks, the number of electrons in the dot remains fixed. The spacing between two successive Coulomb peaks (where resonant tunneling happens) $\Delta V_G$ is determined by $E_{add}$ through a voltage-energy conversion coefficient $\alpha$, known as lever arm, having $E_{add} = \alpha \Delta V_G$ (see Figure 2c). When applying a bias voltage (upper panel of Figure 2c), Coulomb blockade appears in diamond-shaped regions (known as Coulomb diamonds), since $\mu(N)$ also depends on the bias voltage. The insets of Figure 2c mark the level alignment at several typical positions, respectively. The size of the Coulomb diamonds is determined by $E_{add}$ (see Figure 2c), thus containing the information of quantum states.

An alternative way to study the quantum states is via the excited-state spectrum. The discussions above are focused on the ground-state spectrum, which means we only consider the ground state of single-particle level $E_N$. If an excited state (ES) is involved, it will modify $E_N$ thus $\mu(N)$. This results in measurable current change when the related electrochemical potential lies in the bias window as well. The orange line in Figure 2c marks where such peaks can appear. The orange level in the insets of Figure 2c indicates the electrochemical potential where excited state is involved. From such Coulomb diamonds, the energy of excited states $\Delta E(N)$ can be extracted, allowing access to the quantum states.

Therefore, Coulomb diamond measurements are essential to characterize a single quantum dot. It is worth noting that two conditions need to be met to observe Coulomb diamonds. 1) Since $E_C$ (main contribution of $E_{add}$) typically values in the range of a few meV, the temperature needs to be kept low enough to reduce thermal excitation and the bias voltage needs to be kept small enough. 2) The tunneling barriers need to be sufficiently opaque such that the energy uncertainty is much smaller than $E_C$, assuring the electrons locate either in the source, in the drain, or in the dot.

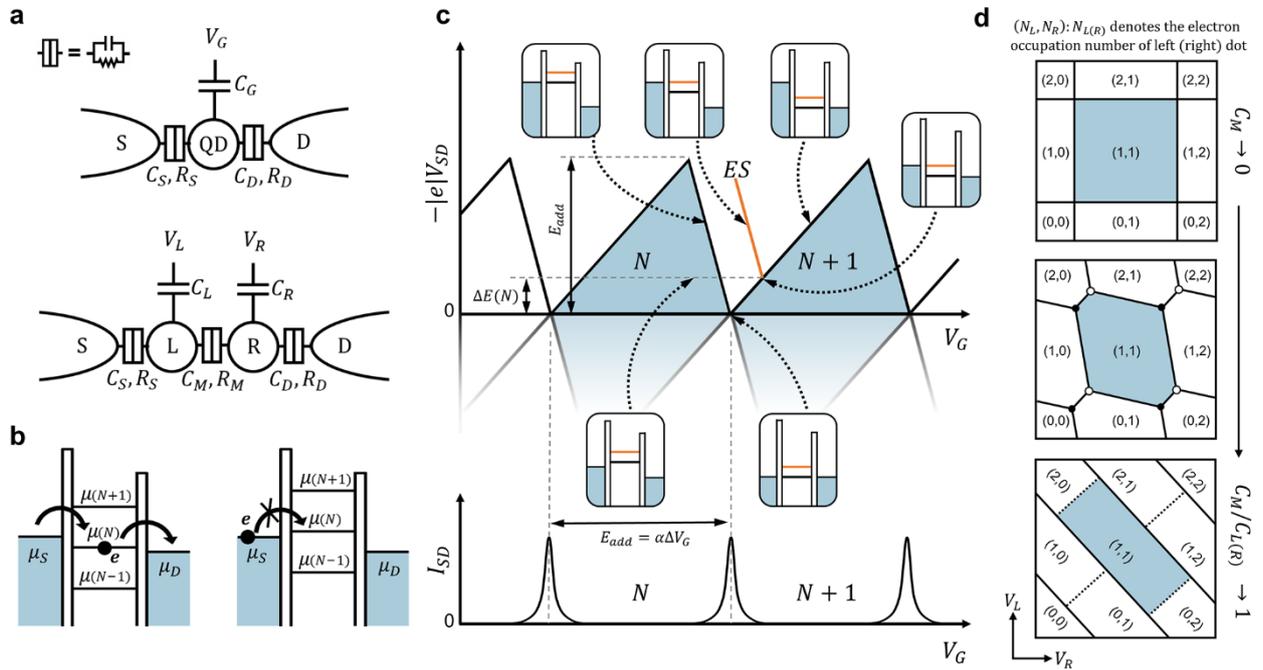

Figure 2. a) Networks of tunnel resistors and capacitors representing single quantum dot (upper panel) and double quantum dot (lower panel). b) Schematic diagrams of the electrochemical potential levels of a single quantum dot. If $\mu(N)$ is tuned into the bias window set by $\mu_S$ and $\mu_D$, the number of electrons can vary between $N$-1 and $N$, resulting in a single-electron tunneling current (left panel). If no level is in the bias window, the electron number is fixed and no current flows, known as Coulomb blockade (right panel). c) Upper panel: schematic plot of source-drain current $I_{SD}$ as a function of $-|e|V_{SD}$

and $V_G$, known as Coulomb diamonds. Insets show the level alignment at several typical positions. The orange level in insets indicates transition related to excited state (ES). Lower panel: $I_{SD}$ as a function of $V_G$, cutting along $V_{SD} = 0$ in the upper panel, known as Coulomb oscillations. Between current peaks, the number of electrons in the dot is fixed. d) Charge stability diagram of a double quantum dot under different tunnel couplings $C_M$. $(N_L, N_R)$ denotes the electron occupation number, in which $N_L$ ($N_R$) represents electron number in the left (right) dot.

For the case of a double quantum dot[75], the charge stability diagram (source-drain current $I_{SD}$ as a function of gate voltages applied to the left ($V_L$) and right dot ($V_R$)) becomes complicated, although the physics is similar. The coupling between the two dots in a double quantum dot can be characterized by interdot capacitance $C_M$. As shown in the upper panel of Figure 2d, for almost decoupled dots ($C_M \to 0$), Coulomb peaks appear along the horizontal and vertical directions in the charge stability diagram, since $V_L$ ($V_R$) only affects the left (right) dot. The number of electrons in the double dot is denoted by ($N_L$, $N_R$). When increasing $C_M$ to an intermedium value, the charge stability diagram exhibits a series of hexagonal-shaped (or honeycomb-shaped) regions, within which ($N_L$, $N_R$) remains fixed (see the middle panel of Figure 2d). For a double dot coupled in series, current can be found when an electron tunnels through both dots sequentially. This condition is met at the so-called triple points (black and white dots in the middle panel of Figure 2d), where transitions between ($N_1$, $N_2$), ($N_1 + 1$, $N_2$), ($N_1$, $N_2 + 1$) occur. When further increasing interdot coupling ($C_M/C_{L(R)} \to 1$), the triple points separate and the honeycomb structure evolves to one set of parallel lines (see the lower panel of Figure 2d). The two dots merge together and behave like a large single dot. Similarly, the quantum states can be electrically accessed and studied through ground-state and excited-state spectra by applying a bias voltage. Thus, gate-controlled quantum dots provide promising platforms for studying the quantum states of electrons at single-particle level. Furthermore, when external fields, such as microwaves or magnetic fields, are employed, different quantum degrees of freedom (including charge, spin, and valley) can be manipulated.

Specifically, for gate-controlled quantum dots based on 2D materials, the most-well studied material system is graphene. Owing to the weak spin-orbit coupling and hyperfine interactions, spins in graphene are expected to have long life time, making graphene a promising candidate for hosting spin qubits[111]. A long coherence time up to 80 μs has been theoretically predicted in graphene quantum dots[112]. However, the lack of a bandgap[44] requires elaborate design on device architectures. Recently, quantum dots based on TMDs also attract research interests[79, 113, 114], due to the semiconducting nature of the materials. Different from graphene, large spin-orbit coupling in TMDs enables faster spin qubit operations[115]. Moreover, both material systems carry valley degrees of freedom (Figure 1b and 1c), which are coupled to their spin degrees of freedom, respectively. Such unique spin-valley couplings allow investigations on valley-related physics and potential applications using valley degree of freedom to encode information[72]. In particular, qubit encoding and operating strategies based on spin[115], valley[116, 117], and spin-valley[113, 118-122] states have been proposed in TMDs, showing great potentials on quantum computation applications based on 2D materials.

## 4 ETCHED QUANTUM DOTS BASED ON 2D MATERIALS

The first 2D material investigated for quantum dot devices is graphene[123-125], shortly after discovering its exotic electronic properties. The gapless electronic band structure makes it challenging to electrostatically confine carriers in graphene. One common way is to geometrically pattern the graphene flake into nanostructure since a sizable gap can be opened when graphene is narrowed down to nanoribbons[126]. Therefore, plasma etching methods[127-132], as well as local oxidation etching using atomic force microscopes[133, 134], are widely used to carve graphene into a quantum dot shape. A scanning force microscope (SFM) image of a typical etched quantum dot device[135] is shown in Figure 3a. The light gray areas are graphene, with their boundaries (white dashed curves) etched using plasma. The size of the quantum dot is around a few tens of nanometers. Two narrow constrictions connect the dot to the reservoirs, forming the tunneling barriers. S, D, and PG denote source, drain, and plunger gate, respectively. Here PG and the conducting channel are made of the same flake of graphene[123, 135-137]. Other designs using adjacent metal gates are also used[138, 139]. Usually, the heavily doped silicon substrate acts as a back gate, separated from the conducting channel with an insulating layer of covered silicon oxide.

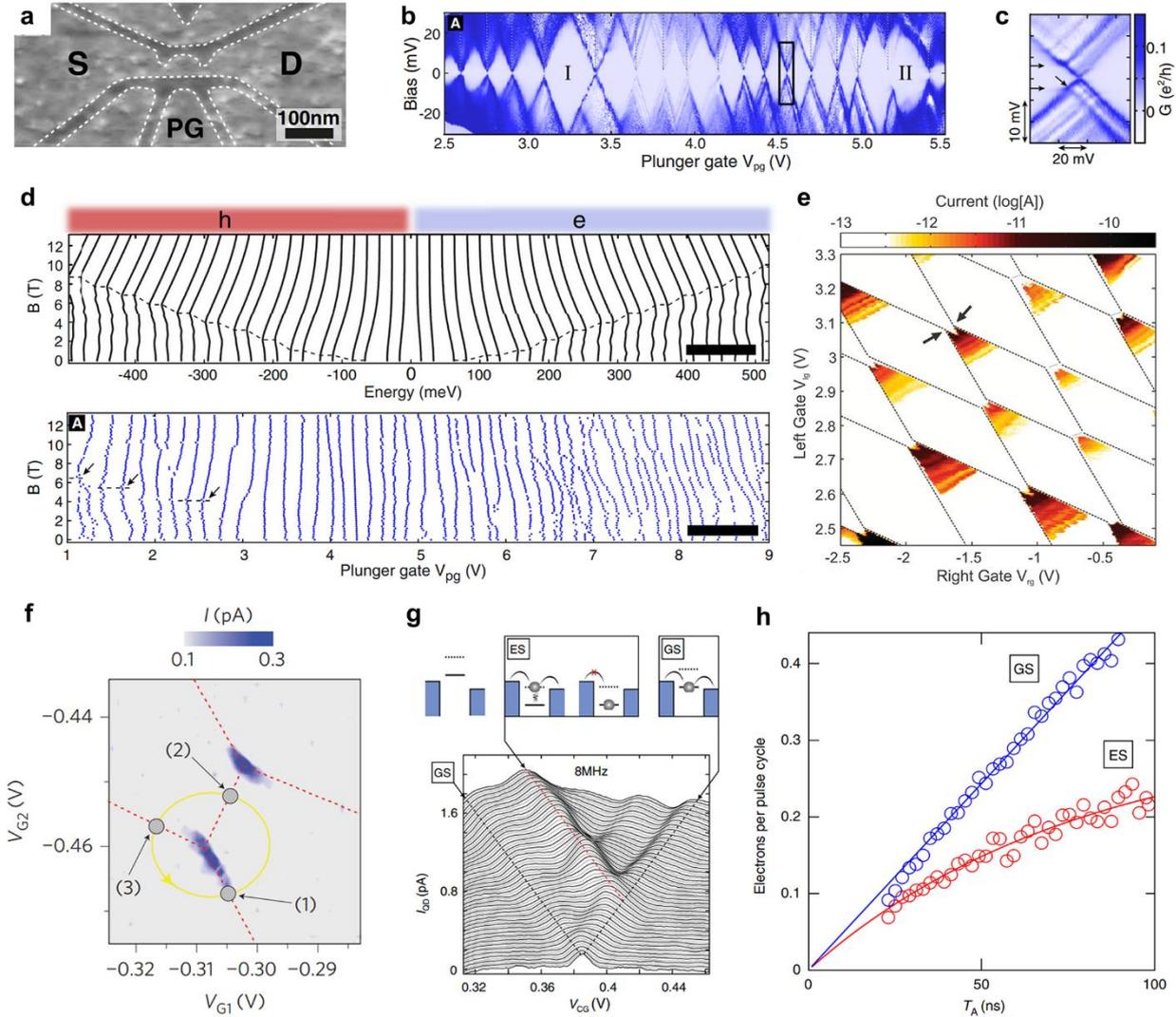

Figure 3. a) SFM image of a typical etched graphene quantum dot device. b) Coulomb diamonds measured in etched graphene quantum dot. c) Zoom-in of the black box in (b), where excited states are clearly resolved. d) Simulation (upper panel) and experimental (lower panel) results of Coulomb peak positions as a function of perpendicular magnetic field, showing a crossover from holes to electrons. e) Hexagonal-shaped charge stability diagram of an etched graphene double quantum dot. f) Source-drain current as a function of voltage applied to gates G1 and G2 of an etched graphene double quantum dot. Charge pumping can be achieved through applying a driving microwave to modulate the gate voltage following the yellow trajectory encircles a triple point. g) Lower panel: Source-drain current through an etched graphene single dot while applying a pulse to the gate CG. Different lines correspond to different amplitudes of the pulse. Increasing

the pulse height, the Coulomb peak splits in two resonances, corresponding to transport through ground state (GS) and excited state (ES), respectively. The corresponding level alignment is illustrated in the upper panel. h) Average number of electrons transferred per cycle through GS and ES as a function of pulse duration $T_A$, from which the charge relaxation time can be extracted through fitting. a-d) Reproduced with permission[135]. Copyright 2009, American Physical Society. e) Reproduced with permission[140]. Copyright 2011, American Chemical Society. f) Reproduced with permission[131]. Copyright 2013, Springer Nature. g, h) Reproduced with permission under the terms of a creative commons CC-BY international license[132]. Copyright 2013, The Authors, published by Springer Nature.

As shown in Figure 3b, Coulomb diamonds have been demonstrated in etched graphene quantum dots[135]. Excited states can be clearly resolved[135] (Figure 3c). Moreover, by applying a magnetic field, spin states can be distinguished. This is achieved by tracing the positions of Coulomb peaks upon varying the magnetic field. Here an in-plane magnetic field is employed to exclude the influence of orbital states. From Zeeman splitting of the spin states, a spin *g* factor of ~2 is extracted[129]. Furthermore, due to the bipolar transport feature of graphene, a change of conducting carriers from electron to hole can be observed in graphene quantum dots[135]. This is demonstrated by measuring the evolution of Coulomb peaks near the charge neutrality point in a perpendicular magnetic field (Figure 3d). When increasing the magnetic field from the low field regime to the Landau regime, different behaviors at electron and hole sides highlight the electron-hole crossover with the $n = 0$ Landau level independent of magnetic field.

Double quantum dots, both coupled in series[127, 137, 138, 140, 141] and in parallel[142], have also been achieved, respectively (Figure 3e). Moreover, when applying a driving microwave to rapidly modulate gate voltage near the triple points, a single electron can be transferred per modulation cycle, demonstrating the phenomenon known as charge pumping (Figure 3f). The driving frequency $f$ determines the rate at which charges are transferred, thus the pumped current is $I = ef$. In graphene, charge pumping at frequencies up to gigahertz is achieved[131].

Time-resolved dynamics of charge carriers have also been investigated in etched graphene quantum dots[132, 143-145]. Employing pulsed-gate spectroscopy, charge relaxation time in graphene quantum dot is measured[132]. As shown in Figure 3g, with a voltage pulse applied to the gate, the excited state is shifted into the bias window. Such an additional tunneling path results in an additional current peak. However, if the electron at the excited state (ES) relaxes to the ground state (GS), such tunneling will be blocked due to energy misalignment, thus reducing the measured current. In this way, by varying the pulse duration, the electron dynamics can be revealed. As shown in Figure 3h, through fitting the saturation behavior of electron number tunneled per cycle as a function of duration time of the pulse, a lower bound of charge relaxation time can be extracted to be 60-100 ns in etched graphene quantum dot[132].

Although substantial experimental progress has been made, it is found difficult to deplete the carriers in the dot to few in this system. Such few-carrier regime is essential for realizing quantum information applications[23]. Also, Coulomb diamonds are usually non-periodic and the barrier transmission is hard to be tuned monotonically by electrostatic gates[77]. These suggest that etched graphene quantum dots are heavily influenced by charge disorders. By employing hBN, a two-dimensional insulating material that dramatically increases the performance of graphene electronic devices[146, 147], as substrate, researchers try to reduce the influence of the disordered potential induced by substrates[148]. However, the improvement is not enough for overcoming the problem. This is because the charge puddles and edge states induced in the etching process result in randomly distributed potentials, which severely influence the quantum dot[77, 149] (Figure 4a). Meanwhile, these localized states can also lead to unintentional multiple dots formation, which is evidenced by transport[150] (Figure 4b) and scanning probe microscopy[151] measurements. Moreover, investigations on potential fluctuations reveal a larger noise level in etched graphene nanoribbons, compared with that of GaAs-based quantum dots. And no obvious difference is found for suspended and unsuspended graphene devices (Figure 4c). This suggests that edge states, rather than the substrate, dominate the larger noise level[152].

To solve this problem, different approaches have been considered. One solution is smoothing the edges. For example, graphene nanoribbons with atomically smooth edges can be achieved through unzipping[153] or squashing[154] a carbon nanotube. This enables tuning the device into few-carrier regime, demonstrating single electron and hole occupation[155]. Such developments provide great potentials for realizing quantum electronic devices based on graphene nanoribbons[156]. Another solution is to induce a band gap in graphene to avoid the etching process, where breakthroughs have been accomplished. Efforts along this direction will be discussed in the next section.

In addition to graphene, etched quantum dots have been demonstrated in other 2D materials. For example, etching $MoS_2$ into nanoribbons can also exhibit Coulomb diamond features[157]. Also, etched quantum dots are demonstrated based on graphene/BN moiré superlattice, where crossover between Coulomb blockade and quantum Hall regimes is investigated[158]. These efforts open the possibilities for realizing quantum dots using the etching process, while technical challenges in achieving full electrostatic control of the dot still remain.

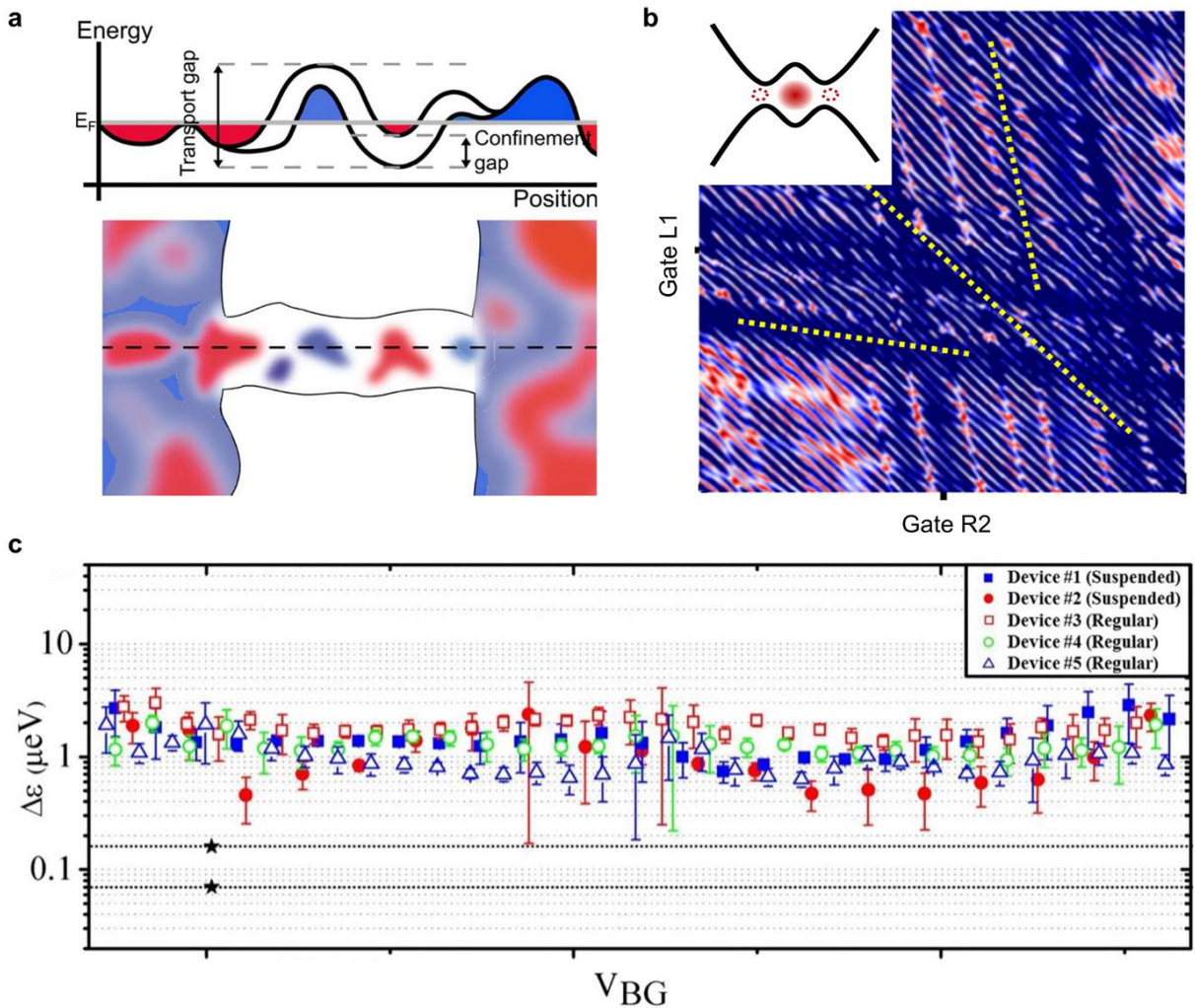

Figure 4. a) Schematic of the randomly distributed potentials in an etched graphene nanoribbon, where electron (blue) and hole (red) puddles form. b) Charge stability diagram of an etched graphene quantum dot, showing three sets of parallel

lines with different slopes (yellow dotted lines). The inset illustrates the quantum dot formation. In addition to the designed single dot (red solid dot), two unintentional dots (open dotted dots) form at the narrow constrictions. c) Potential fluctuations $\Delta\varepsilon$ obtained from suspended and unsuspended (regular) etched graphene nanoribbon devices, showing no obvious difference. Black stars correspond to the noise level of GaAs-based device, which is one order of magnitude smaller. a) Reproduced with permission[149]. Copyright 2010, American Physical Society. b) Reproduced with permission under the terms of a creative commons CC-BY international license[150]. Copyright 2013, The Authors, published by IOP Publishing. c) Reproduced under the terms of a creative commons CC-BY international license[152]. Copyright 2015, The Authors, published by Springer Nature.

## 5  GATE-DEFINED GRAPHENE QUANTUM DOTS

To eliminate the influence of etching process, researchers try to induce a band gap in graphene. Theoretical and experimental studies have shown that by applying a perpendicular electric field to break the inversion symmetry of bilayer graphene, a tunable band gap up to hundreds of meV can be opened[159-162]. Such a band gap engineering approach allows the realization of quantum confinement in biased bilayer graphene via electrostatic gating. Early attempts demonstrate that, through this approach, Coulomb diamonds can be observed, indicating the formation of a quantum dot[163, 164]. However, such early devices have suffered from low pinch-off resistances, which is possibly due to the difficulty in creating a clean and homogeneous band gap. Therefore, none of the devices reaches the few-carrier regime. To overcome this problem, encapsulating bilayer graphene in hBN, as well as edge contact techniques, are employed to minimize the surface roughness and charge disorders induced by the substrate[147]. Moreover, the use of a graphite back gate[165-167] also helps to screen the influence of potential fluctuations induced by the substrate. These technological improvements allow demonstration of quantized conductance and complete current pinch-off upon gating in bilayer graphene quantum point contacts[168-172] (QPCs), which offers possibilities towards high-quality gate-defined graphene quantum dots[173-176].

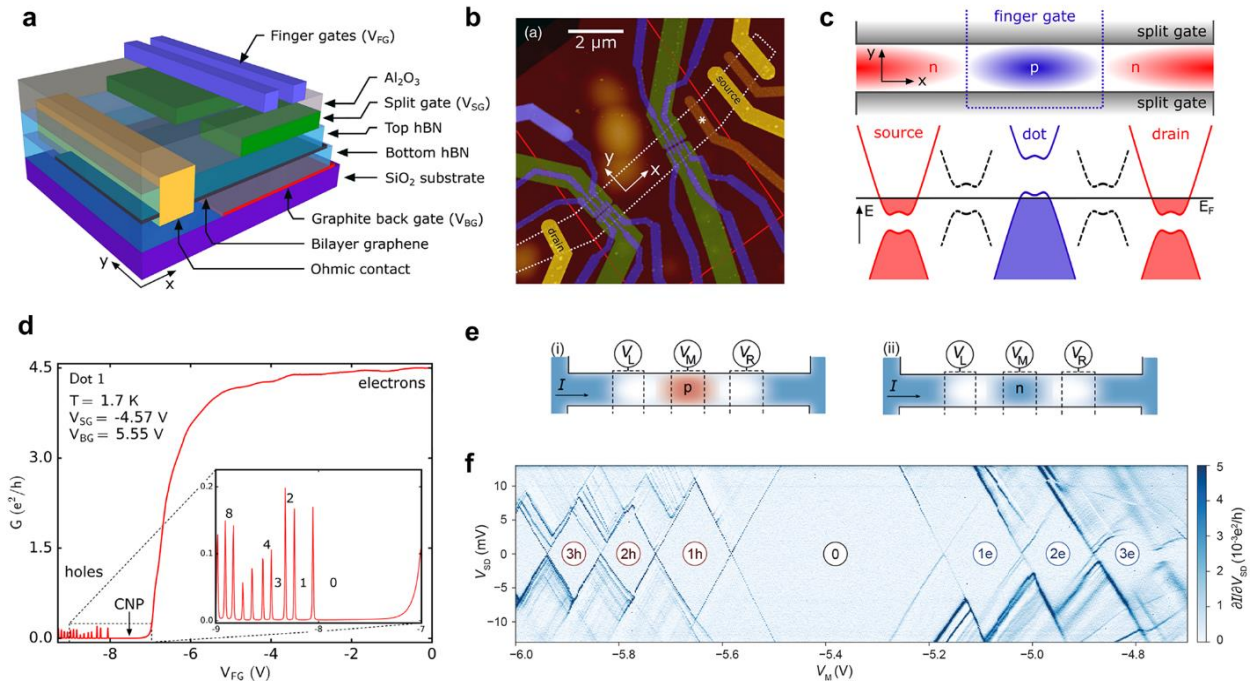

Figure 5. a) Schematic and b) SFM image of a gate-defined bilayer graphene quantum dot device, with parts highlighted

in corresponding colors. c) Schematic of band alignment at different positions along the conducting channel defined by the two split gates. A p-type single dot is confined using the finger gate, with natural p-n junctions (black dashed curves in the lower panel) acting as tunneling barriers. d) Conductance through a gate-defined p-type single quantum dot as a function of finger gate voltage, reaching the few-hole occupation regime. e) Top-view schematic of the conducting channel under different gating configurations, similar to the upper panel of (c). Here, tunneling barriers are formed using gapped (white) regions under finger gates. f) Coulomb diamonds of a gate-defined graphene quantum dot, showing the bipolar operation feature. A transition from a hole single dot to an electron single dot is found, with the corresponding dot formation schematics shown in (e). a-d) Reproduced under the terms of a creative commons CC-BY international license[173]. Copyright 2018, The Authors, published by American Physical Society. e, f) Reproduced with permission[177]. Copyright 2021, American Chemical Society.

Figure 5a shows the schematic of a typical device[173]. The bilayer graphene is sandwiched by top hBN and bottom hBN. Ohmic contacts are achieved using edge contact technique. A graphite flake serves as a back gate. Two split gates are fabricated on the top of the hBN/graphene/hBN heterostructure. Opposite voltages are applied to the split gates and the graphite gate to form a perpendicular electric field to open the band gap. Upon employing proper gate voltages, the Fermi level of the graphene underneath the split gates can be tuned in the band gap, leaving a narrow conducting channel in between. Using an additional layer of finger gates, which are separated from split gates by a layer of insulating $Al_2O_3$, the conductance of the narrow channel can be fully electrostatically pinched off, allowing realization of quantum confinement. The SFM image of the device, consisting of two sets of quantum dots, is shown in Figure 5b, with parts highlighted by corresponding colors in Figure 5a. The white dotted (red solid) lines indicate the boundary of the bilayer graphene flake (graphite back gate).

Using such a device, a single quantum dot can form. Figure 5c shows a configuration of band structure at different positions in the channel, for obtaining a p-type quantum dot. In such configuration, the channel is tuned to operate at n-type, while the area underneath the finger gate is tuned to be p-type. The natural p-n junctions (black dashed curves) in between act as two tunneling barriers, which are responsible for the confinement. As shown in Figure 5d, the few-hole regime, even down to single-hole occupation, can be achieved[173]. It is worth noting that an n-type quantum dot can be similarly achieved if reversed gate voltages are applied, due to the bipolar conducting property of graphene[173]. In addition to forming tunneling barriers using natural p-n junctions, employing additional finger gates to provide insulating areas underneath them (white areas in Figure 5e) can also account for tunneling barriers. This offers higher tunability on the tunneling barriers[178]. When multiple narrow finger gates with small separations are used, the bipolar operation of the quantum dot is achieved[177], as shown in Figure 5f. Typical values of addition energy are ~10 meV, corresponding to the dot size with a radius of ~27 nm. Notably, rich groups of excited states can be observed in the Coulomb diamond measurements. From the energy separation between the first electron and the first hole state, a lower bound of the transport band gap is extracted to be ~30 meV. Double and multiple quantum dots can also be realized using multiple finger gates[179, 180]. Especially, considering different strategies of barriers formation, as well as different polarities of carriers, various quantum dot configurations can be achieved in a single device, by simply changing the gate voltages[178].

In order to obtain better control of the device, further improvement has been made to make the whole area of the quantum dot covered by gate electrodes. This is achieved through using two layers of finger gates[174, 181], which are precisely aligned to each other (Figure 6a and 6b). The high tunability of the device is demonstrated by the charge stability diagram shown in Figure 6c, where rich dot formation configurations can be observed[181] (Figure 6d). For example, in region I, a hole single dot is formed, as evidenced by a set of parallel lines. With increasing voltages applied to gate GL and GR, the hole dot is more and more depleted, and a transition to an electron double dot (hexagonal honeycomb pattern in region II) is observed. When appropriate voltages are applied, an electron-hole double dot can form (region III), which is difficult to realize in conventional bulk materials. Furthermore, in the electron double dot regime (region II in Figure 6c), the interdot coupling can be electrically tuned. As shown in Figure 6e, both capacitive ($E^m$) and tunnel ($t_m$) coupling strengths can be monotonically varied through changing voltage applied to the second-layer gate CG[181]. Such an achievement is extremely challenging in etched graphene quantum dots due to randomly distributed potentials.

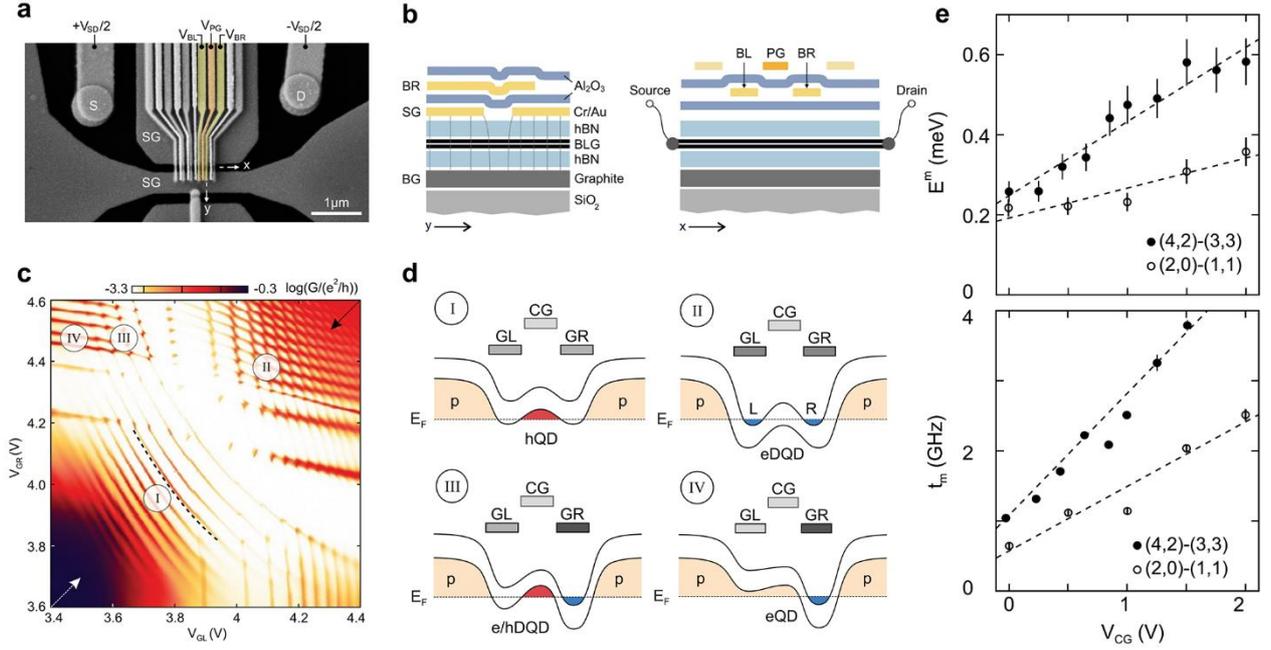

Figure 6. a) False-colored scanning electron microscope (SEM) image of a gate-defined graphene quantum dot, with two layers of finger gates. b) Schematic cross sections of the device along different directions labeled by the white arrows in (a). c) Charge stability diagram of a gate-defined graphene quantum dot, exhibiting different dot formation landscapes upon gating. d) Schematic of band alignment at different regions in (c), with corresponding marks of I-IV, respectively. e) In electron double dot region (II in (c)), measured interdot capacitive coupling energy $E^m$ (upper panel) and tunnel coupling strength $t_m$ (lower panel) as a function of gate voltage applied to gate CG in (d). Both of them can be tuned monotonically with $V_{CG}$. a, b) Reproduced with permission under the terms of a creative commons CC-BY international license[174]. Copyright 2020, American Chemical Society. c-e) Reproduced with permission[181]. Copyright 2021, AIP Publishing.

With such experimental efforts, high-quality quantum confinement with high electrical tunability is realized, which offers an excellent platform for studying spin and valley degrees of freedom in bilayer graphene down to single-particle level. As explained in section 3, $E_{add}$ can be extracted from the spacing between neighboring Coulomb peaks. In few-carrier regime, $E_{add}$ varies as a function of occupation number, due to modifications in single-particle energy spacing $\Delta E$. Subtracting a magnetic field independent charging energy $E_C$, the influence of $\Delta E$ can be better revealed. As shown in Figure 7a, the level spacing is found to bunch in groups of four and increases at the occupation number of four, eight, twelve, and so on, depending on the device[173, 182]. This is due to the two-fold spin and two-fold valley degeneracies in bilayer graphene. Electrons (holes) tend to fill a shell of four states before occupying states of higher energy, exhibiting the level structure, known as shell filling[183]. When the electron occupation number is larger than twelve, an additional three-fold mini-valley degeneracy caused by trigonal warping can be observed[182].

More information about the spin-valley states can be provided via spin and valley Zeeman effects when applying an external magnetic field. As discussed in section 2, valley-contrasting orbital magnetic moments are out-of-plane vectors, thus do not interact with an in-plane magnetic field. Therefore, as shown in Figure 7b, tracing how Coulomb peaks evolve under an in-plane magnetic field gives the spin $g$ factor $g_s$ in bilayer graphene quantum dot to be ~2 (Ref.[173, 182]). Meanwhile, the spin-valley states split in a perpendicular magnetic field due to a combined spin and valley Zeeman effect. Given the spin $g$ factor $g_s$, the valley $g$ factor $g_v$ can be extracted from the single-particle spectrum as a function of perpendicular magnetic field (Figure 7c). The single-particle spectrum is obtained by transferring peak separations from gate voltage to energy using lever arm and subtracting the contribution of charging energy. Typically, $g_v$ is one order of magnitude larger[173, 174, 177] than $g_s$. Moreover, in bilayer graphene quantum dot, changing the dot size by gate voltages can dramatically tune $g_v$ from $29 \pm 2$ to $74 \pm 7$ (Ref.[177]). Such a method using ground-state spectrum of quantum dot has been demonstrated in both hole[177] and electron[173] dots, showing similar features of spin-valley physics.

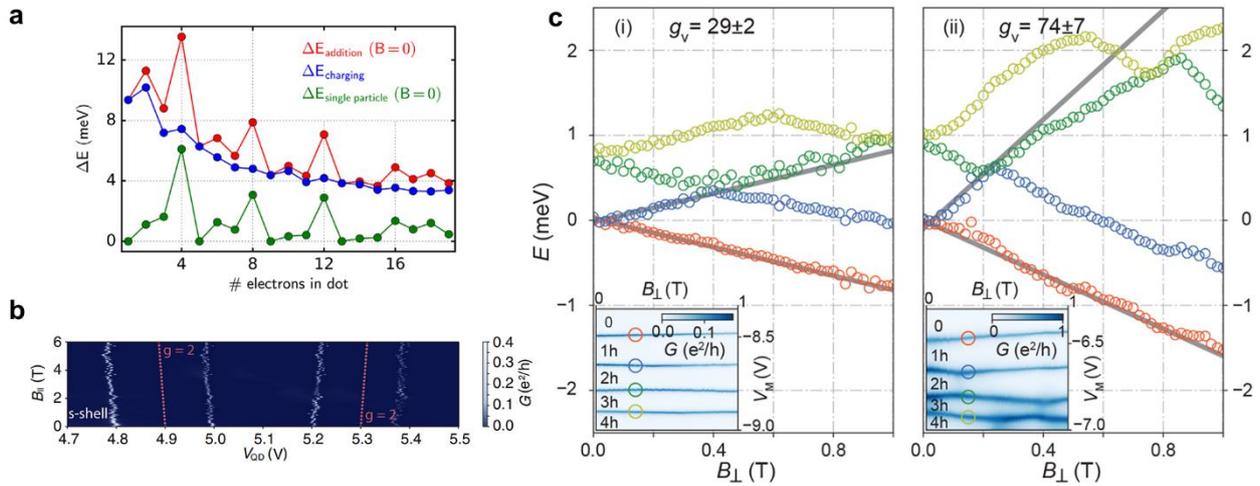

Figure 7. a) Addition, charging, and single-particle energy as a function of electron occupation number in quantum dot. b) Coulomb peaks of the first four electrons as a function of in-plane magnetic field ($B_\parallel$). The red dotted lines correspond to spin Zeeman splitting with the spin $g$ factor $g_s$ of 2. c) Single-particle energy level spectrum of a hole single dot under perpendicular magnetic field ($B_\perp$). The spectrum is extracted from the first four holes Coulomb peaks shown in insets. From linear fittings (gray lines), valley $g$ factors $g_v$ are obtained, which can be electrically tuned under different gating configurations (left and right panels). a) Reproduced under the terms of a creative commons CC-BY international license[173]. Copyright 2018, The Authors, published by American Physical Society. b) Reproduced with permission[182]. Copyright 2021, American Physical Society. c) Reproduced with permission[177]. Copyright 2021, American Chemical Society.

More detailed and quantitative investigations can be carried out when the occupation number is down to only a few. Namely, the one-particle and two-particle states can be experimentally mapped. Figure 8a illustrates the energy dispersion of one-particle states in bilayer graphene quantum dot under in-plane ($B_\parallel$, left) and perpendicular ($B_\perp$, right) magnetic fields[176]. At zero magnetic field, the first four spin-valley states (in the first shell) degenerate into two Kramer's pairs ($K'\uparrow, K\downarrow$) and ($K'\downarrow, K\uparrow$), separated by a gap $\Delta_{SO}$ induced by Kane-Mele type spin-orbit coupling. Here, $K$ ($K'$) is the valley index, while ↑ (↓) denotes the spin orientation. The spin-orbit coupling acts as an effective out-of-plane magnetic field close to the $K$ and $K'$ points, polarizing spin states to have perpendicular directions. An external magnetic field shifts these states with different manners upon directions in which the field is applied. Due to the non-trivial Berry curvature near $K$ and $K'$ points[184, 185], out-of-plane orbital magnetic moments have opposite signs for the two valleys. Therefore, when applying an out-of-plane magnetic field, the four states are shifted by $E(B_\perp) = 1/2(\pm g_s \pm g_v)\mu_B B_\perp$, with $\mu_B$ as the Bohr magnetic moment. Since $g_v$ is one order of magnitude larger than $g_s$, each Kramer's pair splits to a $K$-valley branch that increases with $B_\perp$ and a $K'$-valley branch that decreases with $B_\perp$. When an in-plane magnetic field is applied, it only affects spin degree of freedom. Considering the contribution of an effective magnetic field due to spin-orbit coupling, the energy evolution can be described by $E(B_\parallel) = \pm\sqrt{\Delta_{SO}^2 + (g_s \mu_B B_\parallel)^2}/2$, which gradually evolves to spin Zeeman splitting at high $B_\parallel$.

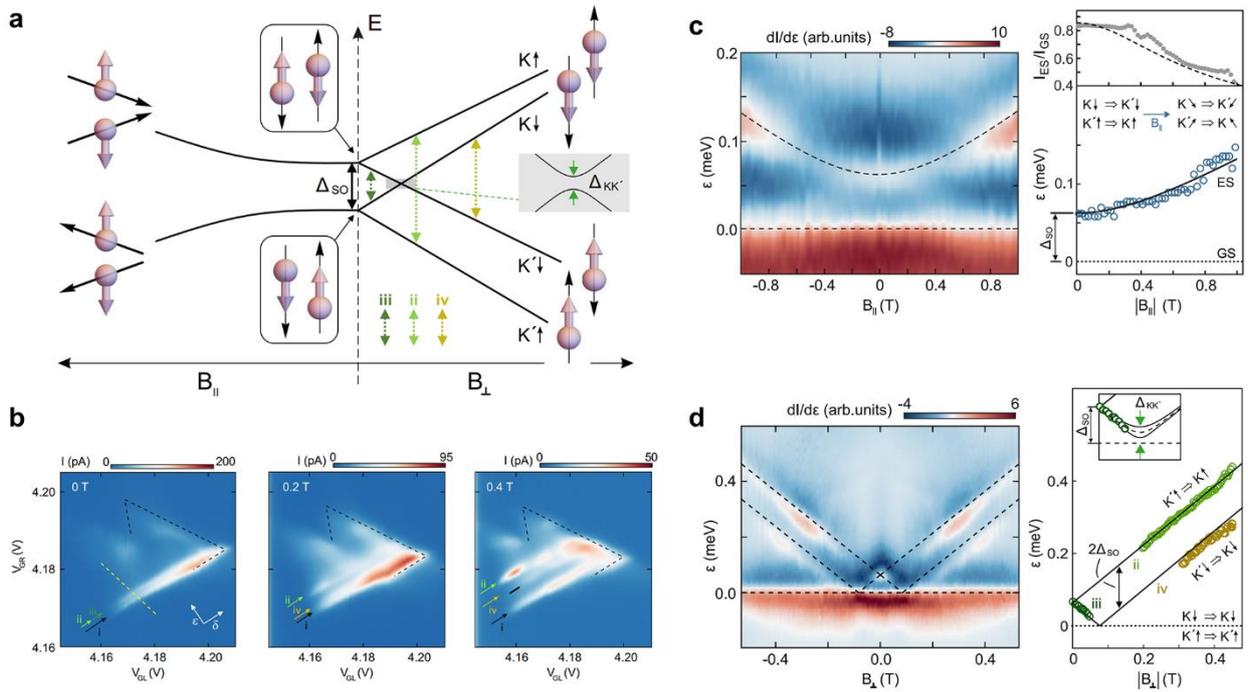

Figure 8. a) Schematic of energy dispersion of one-particle states in bilayer graphene quantum dots as a function of in-plane ($B_\parallel$) and out-of-plane ($B_\perp$) magnetic fields. The four-fold degeneracy is lifted by spin-orbit coupling $\Delta_{SO}$ and separates into two Kramer's pairs (shown in the boxes) at zero magnetic field. b) Charge stability diagrams of a gate-defined graphene double quantum dot under different perpendicular magnetic fields. The applied bias voltage makes triple points evolve to bias triangles, within which transitions related to excited states can be resolved (indicated by corresponding arrows labeled in (a)). c) Left panel: Magneto-transport measurements of the transitions under $B_\parallel$. The dashed lines denote the ground state (horizontal) and excited state (curved) transitions, from which spin-orbit coupling gap $\Delta_{SO}$ can be estimated (right panel). d) Left panel: Magneto-transport measurements of the transitions indicated by the arrows in (b), showing their evolution under $B_\perp$. Right panel: Extracted transition energies (from the left panel) as a function of $B_\perp$, highlighting the transitions (ii), (iii) and (iv) labeled in (a). From the spectrum, disorder-induced valley mixing $\Delta_{KK'}$ (the inset) and $\Delta_{SO}$ can be estimated. a-d) Reproduced under the terms of a creative commons CC-BY international license[176]. Copyright 2021, The Authors, published by Springer Nature.

Such one-particle spectrum can be mapped using the excited-state spectrum of quantum dots. This has been demonstrated in a double quantum dot containing only one electron[176]. The transition between (0, 1) and (1, 0) charge configurations at different magnetic fields is studied. When a finite bias voltage is applied between source and drain, the two triple points in the honeycomb charge stability diagram will develop into two bias triangles (Figure 8b). Inside the triangles, interdot transition lines (peaks in current), which are related to excited states, can be observed along the direction defined as $\delta$ (arrows in Figure 8b). These transition lines correspond to transitions between different spin-valley states of each dot, with corresponding marks ii to iv labeled in Figure 8a. (At the transition line of i, electron transports through the ground state of each dot). Therefore, their evolution under magnetic fields helps in revealing the one-particle spectrum. From the dependency on in-plane (Figure 8c) and out-of-plane (Figure 8d) magnetic fields, the values of $g_s$~2, $g_v$~15, and $\Delta_{SO}$~62 μeV are estimated. Furthermore, the magneto-transport spectroscopy also gives an upper limit of the disorder-induced valley mixing $\Delta_{KK'}$~20 μeV (see inset of Figure 8a and 8d).

In addition, two-particle states in bilayer graphene quantum dots[186], which are more complicated, have also been studied. For both spin and valley wavefunction components, there are three symmetric (triplet state $|T_{S,V}^{0,\pm}\rangle$) and one antisymmetric (singlet state $|S_{S,V}\rangle$) two-particle states, where $S$ ($V$) indicates spin (valley) degree of freedom. Therefore, there are 16 combinations of these components, which can be divided into 6 antisymmetric ($|S_V\rangle|T_S^{0,\pm}\rangle$, $|T_V^{0,\pm}\rangle|S_S\rangle$) and 10 symmetric components ($|T_V^{0,\pm}\rangle|T_S^{0,\pm}\rangle$, $|S_V\rangle|S_S\rangle$). In order to maintain an antisymmetric total wavefunction, antisymmetric (symmetric) spin-valley components need to be combined with symmetric (antisymmetric) orbital components, which results in bunching into two major groups in energy at zero magnetic field. Evolution of these states upon external magnetic fields allows mapping the energy level spectrum, since spin and valley components behave differently under magnetic field with different directions.

Figure 9a shows the energy level of the lowest five states[175] as a function of $B_\parallel$ and $B_\perp$. At zero magnetic field, a three-fold degenerate valley-singlet and spin-triplet state ($|S_V\rangle|T_S^-\rangle$, $|S_V\rangle|T_S^0\rangle$, $|S_V\rangle|T_S^+\rangle$) is separated with a two-fold degenerate valley-triplet and spin-singlet state ($|T_V^-\rangle|S_S\rangle$, $|T_V^+\rangle|S_S\rangle$) by energy spacing of $\Delta E_{Exch}$ ($|T_V^0\rangle|S_S\rangle$ has a higher energy and is not shown in the figure). When applying an external magnetic field, the valley triplets $|T_V^\pm\rangle$ split under $B_\perp$ with $g_v$, while the spin triplets $|T_S^{0,\pm}\rangle$ split under both $B_\perp$ and $B_\parallel$ with $g_s$.

Note that the ground state of two-particle states at zero magnetic field is a valley-singlet and spin-triplet state ($|S_V\rangle|T_S^{0,\pm}\rangle$). This unique property has been demonstrated in experiments through the excited-state spectrum of a hole dot[187]. As indicated by arrows in Figure 9b, transition lines corresponding to excited states can be observed in the Coulomb diamond measurements. Evolution of the energy level spacing $\Delta E$ upon magnetic field can be used to reveal the two-particle states. As shown in the left panel of Figure 9c, the two-hole ground state splits into three components with increasing $B_\parallel$, suggesting the ground state is spin triplets. When further tilting the orientation of magnetic field (Θ) while keeping the value of total field unchanged (right panel of Figure 9c), the induced out-of-plane field component leads to valley splitting with spin states unaffected. As shown in Figure 9c, the results indicate that the ground state at zero magnetic field is a valley-singlet and spin-triplet state. From the experiment, the exchange energy $\Delta E_{Exch}$ is estimated to be 0.35 meV. Recently, further experiments provide a more complete mapping of the whole 16 two-particle states using similar excited-state spectrum of an electron dot. When applying a perpendicular magnetic field, a transition of ground-state, gradually from $|S_V\rangle|T_S^{0,\pm}\rangle$ to $|T_V^-\rangle|S_S\rangle$ then to $|T_V^-\rangle|T_S^{0,\pm}\rangle$, has been demonstrated[188].

Another experimental method for studying spin-valley states in gate-defined graphene quantum dots is using Kondo effect. Similar to Kondo effect in semiconductors, a quantum dot with its net spin can be regarded as localized spin and its nearby leads act as surrounding Fermi sea. Formation of Kondo state is evidenced in the experiment by a current peak at zero bias inside the Coulomb diamond, with peak height decreasing when increasing the temperature[189, 190]. The exotic spin-valley feature of bilayer graphene allows observation of the mixing of the spin and valley Kondo effects in a single dot at few-hole regime (Figure 9d), from which both one-particle and two-particle level spectra can be examined[175]. This can be done through tracing the Kondo resonance's evolution at different occupation numbers, as a function of $B_\parallel$ and $B_\perp$, respectively. As shown in Figure 9e, taking two-particle spectrum as an example, the conductance peaks, with corresponding transitions labeled by α, β, γ and ε in Figure 9a, reveal the energy level splitting scheme in magnetic field illustrated in Figure 9a. Note that transition labeled by β is absent in the experiment since it requires spin flips of both holes.

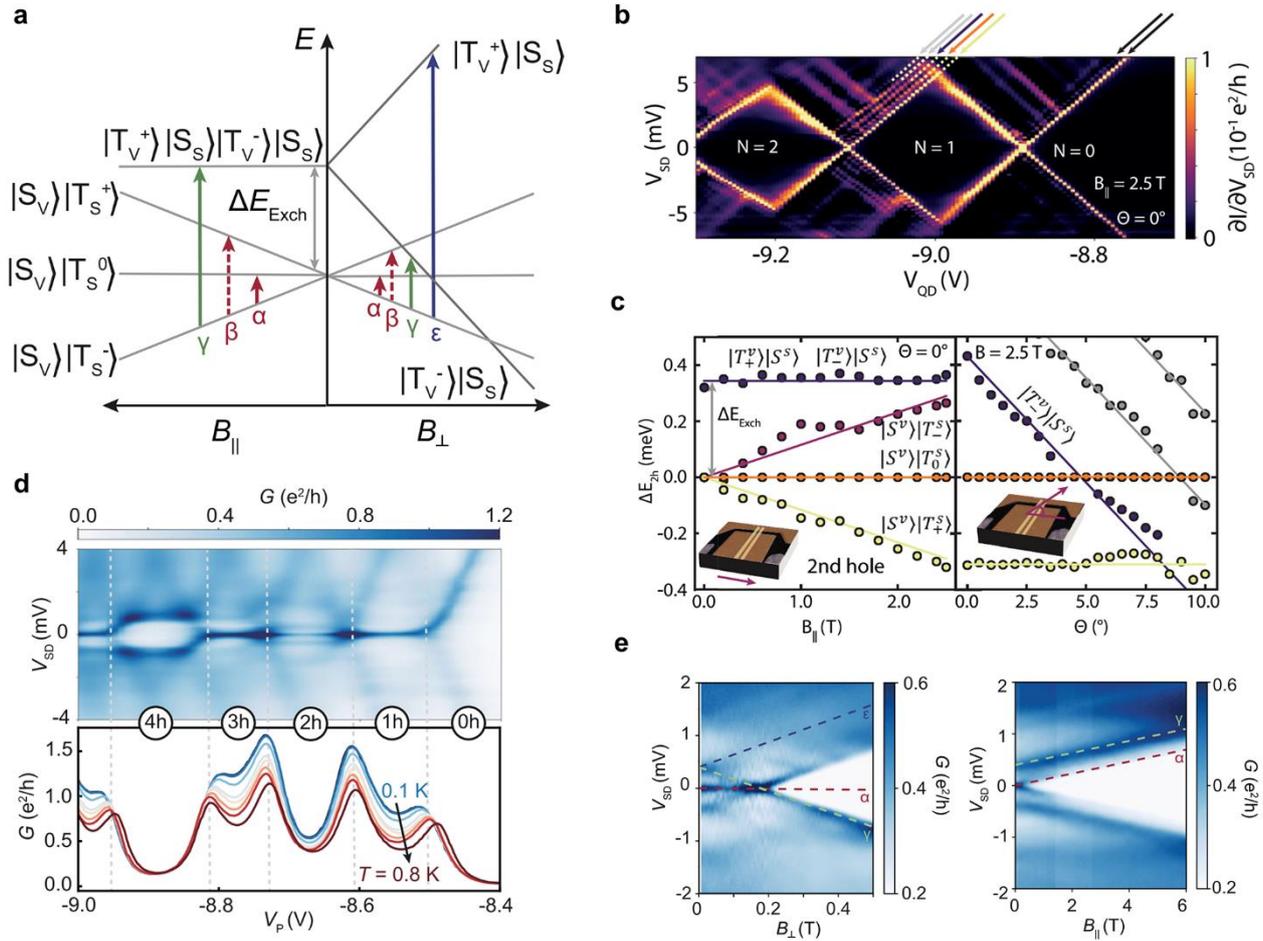

Figure 9. a) Schematic of energy dispersion of two-particle states in bilayer graphene quantum dots as a function of in-plane ($B_\parallel$) and out-of-plane ($B_\perp$) magnetic fields. $|S\rangle$ and $|T^{0,\pm}\rangle$ denote singlet and triplet states, respectively. $S$ and $V$ in subscripts represent spin and valley index, respectively. b) Coulomb diamonds of a gate-defined graphene hole single dot under in-plane magnetic field, in which excited states can be resolved (marked by colored arrows). c) Left panel: Extracted excited state energies as a function of in-plane magnetic field. Right panel: Evolution of excited state energies when titling magnetic field to increase perpendicular field component. d) Upper panel: Coulomb diamonds obtained from a gate-defined graphene hole single dot, showing occupation of first four holes. The conductance peaks at zero bias voltage and their temperature dependence (lower panel) provide experimental evidence of Kondo resonances. e) Kondo resonances under perpendicular (left panel) and in-plane (right panel) magnetic fields. The dashed lines label transitions with corresponding marks in (a). a, d, e) Reproduced under the terms of a creative commons CC-BY international license[175]. Copyright 2021, The Authors, published by Springer Nature. b, c) Reproduced with permission[187]. Copyright 2019, American Physical Society.

Besides mapping the spin-valley states, Pauli blockade is demonstrated recently in gate-defined graphene quantum dot[191]. Tunneling through particular spin (valley) states can be blocked upon gate and magnetic field tuning, due to Pauli exclusion principle of spin (valley) degree of freedom. As a consequence, a suppression of transport current is observed, typically at bias triangles of a double dot with few carriers, resulting in spin (valley) blockade. Furthermore, dynamics of excited states are also studied using pulsed-gate spectroscopy[192-194]. A spin relaxation time exceeding 200 μs has been extracted[193].

In addition to using an electric field to induce a band gap in bilayer graphene, other strategies are also considered and experimentally demonstrated to form quantum dots in graphene without employing the etching process. For example, gate-tunable quantum confinement has been achieved using quantum Hall gap[164] or local strain[195]. Also, edge-free graphene quantum dots can be induced by the tip of a scanning tunneling microscope[196-198].

# 6  GATE-DEFINED QUANTUM DOTS BASED ON 2D SEMICONDUCTORS

Another way of avoiding the etching process is to use 2D semiconductors, which naturally have bandgaps. Following this direction, gate-defined quantum dots have been studied in 2D semiconductors, firstly in representative materials of TMDs[199]. TMDs, such as $MoS_2$, $WS_2$, $WSe_2$, and $MoSe_2$, naturally have a band gap[200] of 1 to 2 eV, allowing demonstration of field-effect transistors (FETs) with a high on-off ratio[201]. QPCs based on TMDs have also been realized, in which the quantized conductance is observed and can be electrostatically pinched off[202-205]. Moreover, the coupled spin-valley degrees of freedom in TMDs provide possibilities for encoding information as well[73]. Besides the common K valleys, which are similar to those studied in graphene, carriers from other valleys can also dominate the transport behavior in TMDs, due to the unique band structure and its layer-sensitive feature. In particular, flavor quantum dots have been proposed based on electrons from the triply degenerate Q valleys in few-layer TMDs[206, 207]. Furthermore, from the view of spin qubits, the large spin-orbit coupling in TMDs enables fast qubit operations upon electrical control. Theoretical studies predict that Rabi oscillations with frequency as fast as 250 MHz can be achieved in TMDs quantum dots[115].

Due to the natural bandgap, a more straightforward design of gate architecture is used to form quantum dots in TMDs, compared with its counterpart in bilayer graphene. Controlled confinements of individual charge carriers are firstly demonstrated in single quantum dots. A typical device[199] is shown in Figure 10a, where few-layer $WSe_2$ is electrically accessed through metal source and drain. A heavily-doped Si substrate acts as a back gate (BG) to globally dope the $WSe_2$ layer to conduct. Four top gates (LB, PG, RB, and MG), separated from $WSe_2$ by a layer of insulating $Al_2O_3$, are responsible for depleting the underneath regions, forming the quantum dot. Figure 10b shows transport measurements of the device, where consecutive Coulomb diamonds with uniform sizes are observed[199]. Similar architecture is also applied to realize quantum dots on $WS_2$.[208, 209] Further progress is made to improve the quality of these early devices by employing hBN encapsulation. Figure 10c shows the schematic of a few-layer $MoS_2$ device with hBN encapsulation[210]. The sandwiched heterostructure is transferred onto a substrate with pre-patterned bottom gates for confining the dot. Note that edge contact to TMDs is technically challenging. Thus, the $MoS_2$ flake is bridged by graphene flakes for electrical contact to metal contact gates. With similar device architectures, single quantum dots have been demonstrated in few-layer $MoS_2$,[210] bilayer $WSe_2$,[211] and monolayer $WSe_2$.[211, 212] Notably, the number of holes confined in the single dot is estimated to be in the range of 10-20, approaching the few-carrier regime. As a consequence, excited states can be observed in the Coulomb diamond measurements[211] (arrows in Figure 10d).

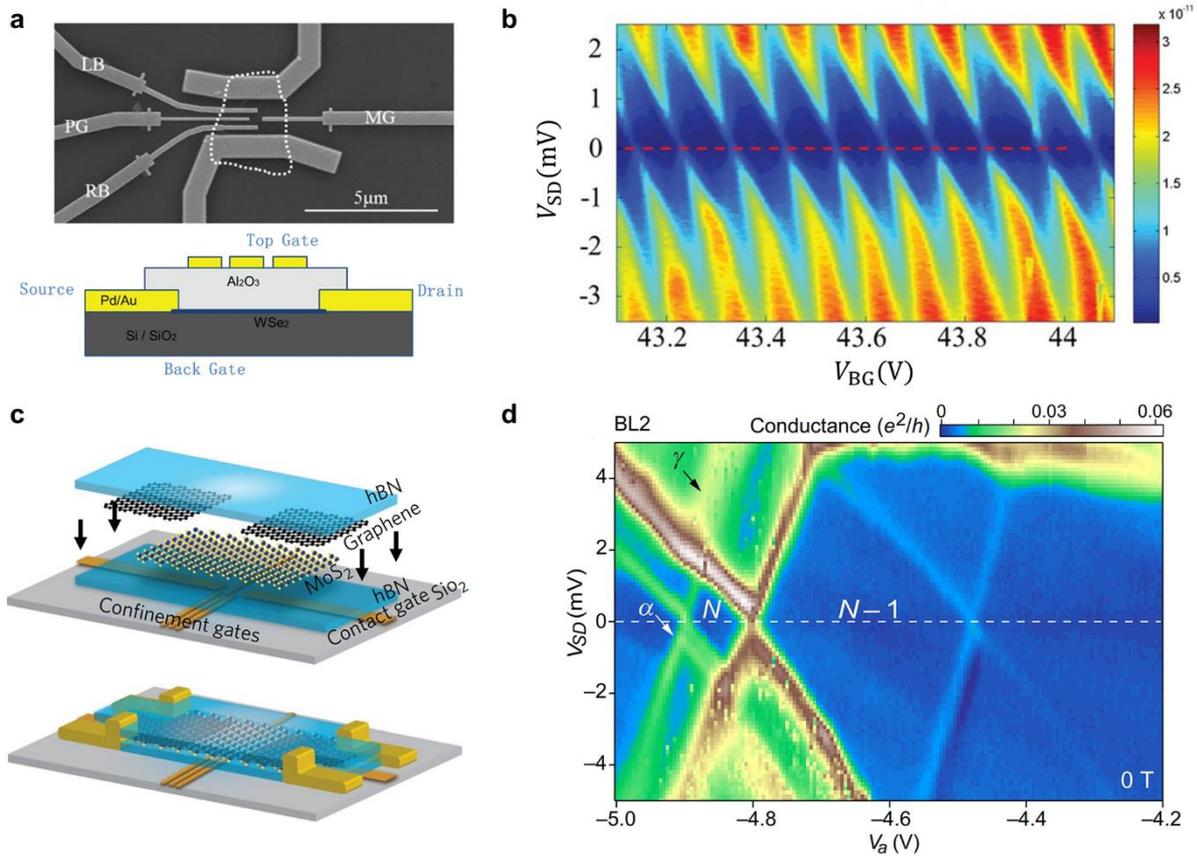

Figure 10. a) SEM image (upper panel) and cross-sectional schematic (lower panel) of a gate-defined $WSe_2$ quantum dot. The white dotted line labels the $WSe_2$ flake. b) Coulomb diamonds obtained from the device shown in (a). c) Schematic of a gate-defined $MoS_2$ quantum dot device with hBN encapsulation. The metal source/drain contacts are connected to $MoS_2$ channel via two graphene flakes. d) Coulomb diamonds of a bilayer $WSe_2$ quantum dot, showing evidence of excited states (indicated by the arrows). a, b) Reproduced with permission[199]. Copyright 2015, Royal Society of Chemistry. c) Reproduced with permission[210]. Copyright 2018, Springer Nature. d) Reproduced with permission[211]. Copyright 2020, American Physical Society.

Besides single dots, double quantum dots are also realized in TMDs, using hBN encapsulation[213, 214]. However, the electrical contact strategies are different, including direct top contact[213] (Figure 11a) and via-hole contact[214] methods (Figure 11b). A typical gate design is shown in Figure 11c, with S and D denoting source and drain contacts, respectively. Six gates (LB, LP, DM, RP, RB, and UM) define the confining potential for the double quantum dot[213]. Charge stability diagram of the device is shown in Figure 11d, where a hexagonal honeycomb pattern is observed, suggesting the formation of the double dot. Through tuning voltage applied to the gate DM, the pattern gradually evolves to a set of parallel lines (insets of Figure 11e), indicating the transition from double dot to merge into a larger single dot (Figure 11f). Such monotonic tuning of interdot coupling (Figure 11e) demonstrates high tunability of gate-confined quantum dot based on TMDs.

It is worth noting that quantum dots based on TMDs have usually suffered from charge impurities[208, 215, 216], possibly due to the large effective mass and relatively low mobility. These charge impurities are usually localized and have small sizes, which can be experimentally evidenced from a large extracted $E_{add}$. [208, 215] Also, Coulomb blockade weak antilocalization has been demonstrated at low-density regime, implying a combined influence of spin-orbit coupling and short-range disorders[213].

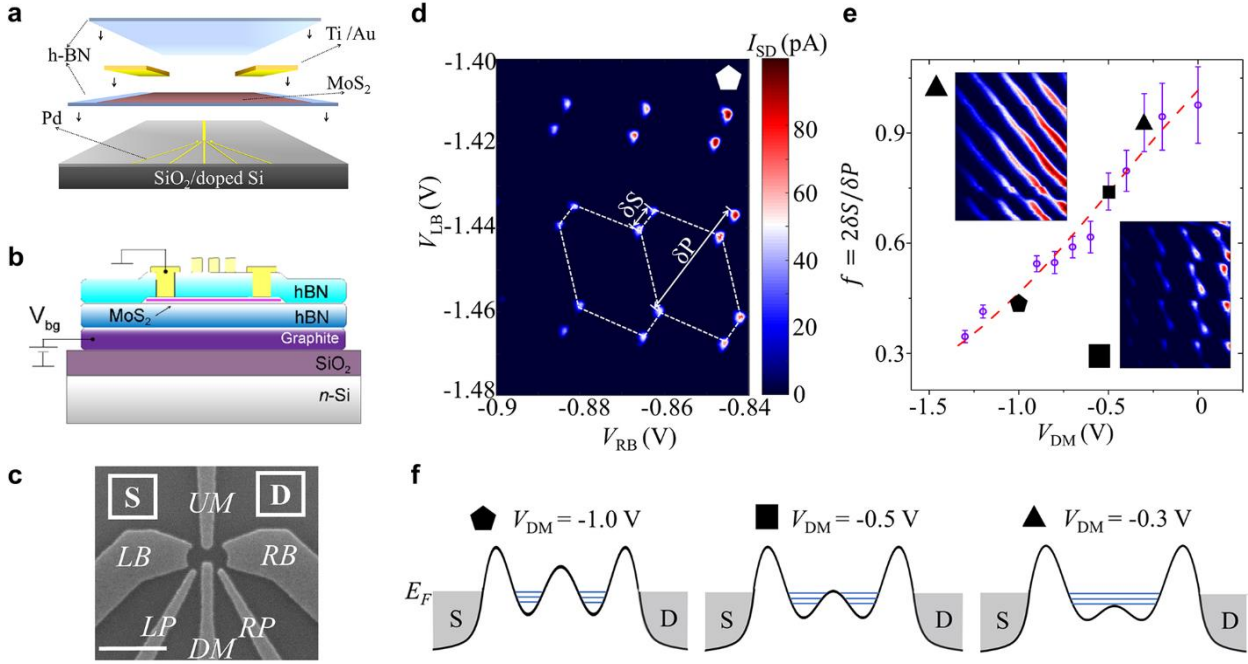

Figure 11. a, b) Schematic of gate-defined $MoS_2$ double quantum dots with hBN encapsulations. The electrical contacts are achieved using (a) direct top contact and (b) via-hole contact method, respectively. c) SEM image of the gate design of the device shown in (a). d) Charge stability diagram of the device shown in (a), exhibiting hexagonal-shaped feature. e) $f = 2\delta S/\delta P$ increases monotonically with increased gate voltage $V_{DM}$, suggesting a transition from two weakly-coupled dots to a merged single dot. Insets show charge stability diagram obtained at different $V_{DM}$ with corresponding marks. f) Schematic of energy landscapes, illustrating the transition from double dot to single dot regime. a, c-f) Reproduced under the terms of a creative commons CC-BY-NC international license[213]. Copyright 2017, The Authors, published by American Association for the Advancement of Science. b) Reproduced with permission[214]. Copyright 2018, AIP Publishing.

Another successful example for demonstrating gate-defined quantum dot is achieved using InSe,[217] a 2D semiconducting material with high mobility[218]. Since InSe degrades under ambitious conditions, encapsulation it using hBN in an inert environment is essential to avoid exposing the flake to the air[218]. A similar graphene-mediated electrical contacting strategy is used, as shown in Figure 12a. Coulomb diamonds have been observed in such InSe-based van der Waals heterostructures (Figure 12b), showing the capability of quantum dot formation[217].

In addition to quantum dots, Coulomb diamonds have also been observed in single-electron transistors (SETs) based on 2D semiconducting materials[219]. Recently, electrostatic gate confinement has been realized in SETs based on twisted graphene systems[220, 221]. In specific, upon gating, magic-angle twisted bilayer graphene (MATBG) can be tuned into different phases of superconducting, metallic, and insulating[61, 62]. Employing devices with multiple gate electrodes, tunneling barriers can be formed using the insulating states (Figure 12c). Therefore, electrons can tunnel through these barriers individually, exhibiting features of Coulomb diamonds[220] (Figure 12d).

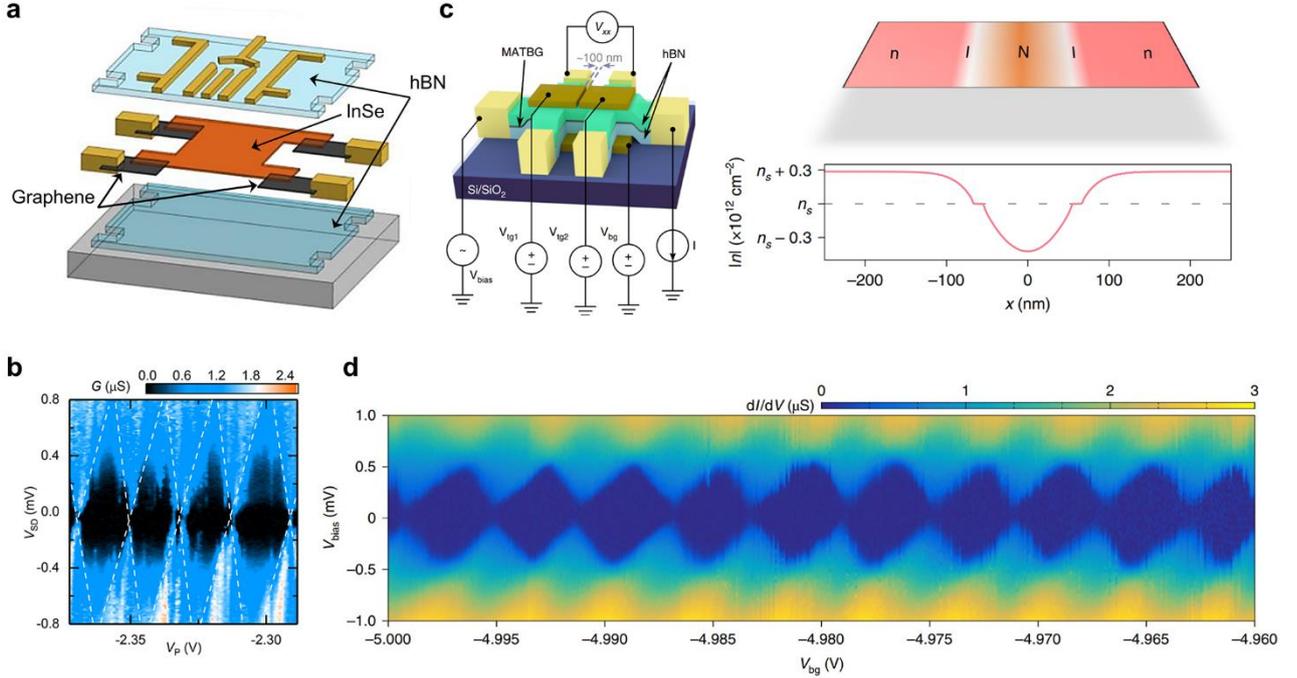

Figure 12. a) Schematic of a gate-defined quantum dot device in InSe-based van der Waals heterostructures. b) Measured Coulomb diamonds of device shown in (a). c) Left panel: Schematic of a multi-gate MATBG device. Right panel: Gating scheme for an electrostatically defined SET, with insulating areas (labeled by I) acting as tunneling barriers. d) Coulomb diamonds of the MATBG device shown in (c), suggesting the formation of a SET. a, b) Reproduced with permission[217]. Copyright 2018, American Chemical Society. c, d) Reproduced with permission[220]. Copyright 2021, Springer Nature.

## 7 CHALLENGES AND OPPORTUNITIES

Among gate-controlled quantum dots based on different 2D materials, devices of biased bilayer graphene lead the race towards realizing qubits at present. However, there are still some obstacles yet to be overcome. To name a few, the first one is the readout scheme. The most widely used scheme in works discussed above is based on direct transport measurement[173-176], in which a complete tunneling process through both the source-dot and dot-drain barriers needs to be accomplished to form a transport current. However, this scheme can be extremely challenging under circumstances where dynamics of spin/valley states needs to be resolved. For example, the tunneling barrier needs to be sufficiently opaque in the measurement of spin relaxation time, with tunneling rates having typical values of several MHz. This results in a transport current below 1 pA, which is very difficult to be precisely measured[192-194]. Meanwhile, particular electron state manipulations require that the electron tunneling can only happen through single-side barrier[222]. In these cases, no transport current can be detected. One possible solution is using SETs or QPCs nearby as charge detectors[128, 223] (Figure 13a). The detectors are capacitively coupled to quantum dot and sensitive to the change of charge environment in the dot. Thus, when a tunneling event happens in the dot, it will be reflected by a current jump/peak of the detector, even when direct transport current through the dot is no longer measurable (Figure 13b). Such a charging detecting scheme has been demonstrated in both etched[128, 138, 143, 145, 224] and gate-defined[194, 225] graphene quantum dots. Another possible solution to the readout problem is using the radio-frequency reflectometry technique[226]. This method detects the change in quantum capacitance when a tunneling event happens and has been demonstrated in gate-defined bilayer graphene quantum dot[227] (Figure 13c). Opportunities can be expected on investigations of time-resolved dynamics[145, 194, 228-232] in gate-defined graphene quantum dots using these readout schemes.

Despite the rapid progress in mapping the spin-valley states in gate-controlled quantum dot based on bilayer graphene, how to dynamically control these states is another aspect that needs to be addressed. Careful design and precise control over microwave and pulse sequences are essential for encoding and manipulating information based on these different quantum degrees of freedom[34, 42, 233, 234]. Thus, the devices' response to microwaves and pulse sequences needs to be investigated to reveal the coherency of the system. Although charge pumping[131] and relaxation time measurements[132, 192-194] have been realized, other typical microwave manipulation experiments, which are widely studied in quantum dots based on other material systems, such as photon-assisted tunneling[235-238], and controlled rotations on the Bloch sphere[239-248], are still in lack of demonstration in 2D material systems.

Another challenge is the integration of multiple quantum dots. This can be realized not only by constructing a dot array using neighboring interactions[249-251], but also through hybrid architectures[19, 38, 40, 252, 253]. A typical system is integrating quantum dots with a microwave cavity[139] (Figure 13d), in which photons can interact with electrons in the quantum dot. It allows an accurate readout of the electron state through the response of the cavity[139, 254-256] (Figure 13e). Moreover, the microwave cavity can act as a bridge to realize long-range interactions between different sets of quantum dots[40, 257, 258]. This architecture has been demonstrated in etched graphene quantum dot system[139, 257, 259], where a four-fold periodicity in dephasing rates is extracted[139]. Next step is expected to achieve strong coupling between photons and different quantum degrees of freedom, which has been successfully realized in quantum dots based on other material systems recently[22].

Furthermore, gate-controlled quantum dots can also be realized using suspended graphene flakes[164] (Figure 13f). In such suspended devices, the graphene flakes can oscillate, serving as mechanical resonators simultaneously[260-262]. Mechanical excitations of the mechanical resonators are regarded as vibrational phonons, which are promising candidates for information processing[263-267]. Previous experiments have shown excellent coherency of the vibrational phonons host in graphene mechanical resonators[268, 269]. Thus, such hybrid architectures consisting of graphene quantum dots and mechanical resonators[270] may present a novel way for phonon-based quantum dot integrations and applications, which have already been widely studied in its 1D counterpart of carbon nanotube[271-275].

For TMD gate-defined quantum dots, there are more basic issues that need to be addressed. For example, the quality of the flake, as well as the interface, needs to be further improved to reduce the influence of the commonly observed defects and disorders[107, 276, 277]. Another important aspect is the electrical contact to TMDs, which has received lots of attention in TMD FET studies in the past decades[278, 279]. Recently, several methods, such as metal contact transferring[280], van der Waals contacting[281], and suppression of metal-induced gap states[282], show excellent contact quality at room temperature and even down to 77 K. However, their performances at extremely low temperatures (below 4 K), and compatibilities with quantum dot devices are yet to be investigated. Moreover, the gate electrode design is also needed to be reconsidered. Due to a larger effective mass[74], smaller confinement is needed in TMDs. Therefore, overlapping gate architecture, which is widely used in silicon-based quantum dots[40, 244, 245], may be helpful for lowering the size of the dot to reach the last carrier occupation.

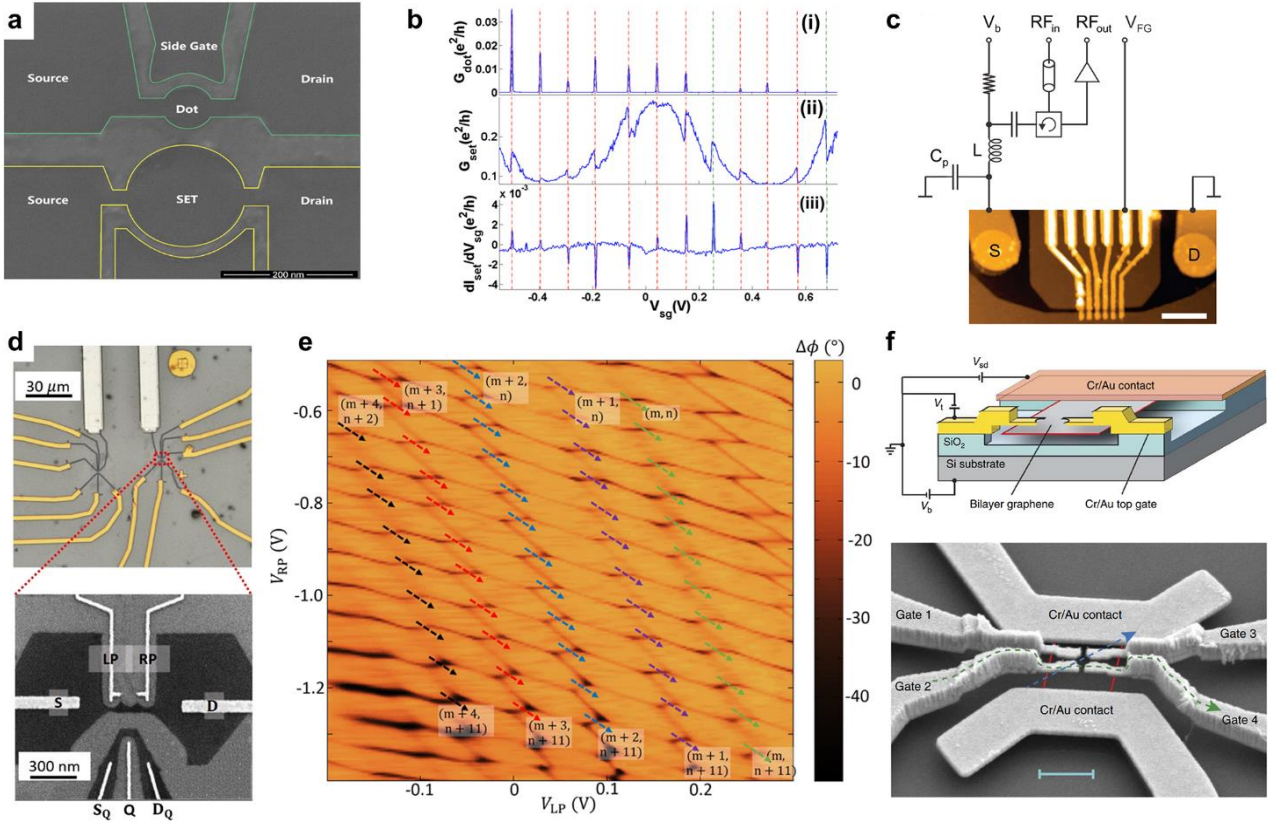

Figure 13. a) SEM image of an etched graphene quantum dot with a SET nearby serving as a charge detector. b) Transport measurements of the device shown in (a). Even the direct transport current through the dot (i) is no longer measurable, SET transport (ii) and SET modulation measurements (iii) can still detect single-electron tunneling events. c) Schematic of the measurement setup for dispersive sensing a gate-defined graphene quantum dot. d) An etched graphene quantum dot integrated with an on-chip superconducting microwave resonator. e) Charge stability diagram of the device shown in (d), measured by phase response $\Delta\phi$ of the reflection line resonator. f) Schematic (upper panel) and SEM image (lower panel) of a gate-defined quantum dot based on suspended bilayer graphene. a, b) Reproduced with permission[128]. Copyright 2010, AIP Publishing. c) Reproduced with permission[227]. Copyright 2021, AIP Publishing. d, e) Reproduced with permission[139]. Copyright 2015, American Physical Society. f) Reproduced with permission[164]. Copyright 2012, Springer Nature.

Finally, we would like to briefly compare gate-controlled quantum dots based on 2D materials with those on conventional semiconductors (such as Si, Ge), from the view of constructing qubits. Taking graphene as an example, significant efforts have been devoted to improving the mobility of graphene to ~$10^6$ cm$^2$/Vs in the past decades[146, 147, 283-285], suggesting the high quality of graphene crystals as well as their interface to insulating layers. The low density of disorders is essential for realizing qubits. Moreover, due to the weak spin-orbit coupling and hyperfine interaction, graphene is expected to be a promising material for hosting spin qubits[111]. The long spin relaxation time in graphene quantum dots reported recently[192-194] provides great potentials along this approach.

Comparing with conventional semiconductors, a sharp but exciting difference is the unique spin-valley physics in 2D materials. In conventional semiconductors quantum dot systems, the most well-studied qubit encoding strategy is based on spin degree of freedom[286, 287]. The valley degree of freedom is usually recognized as an undesirable constraint since it will affect the spin qubit[232, 288-291]. Although the valley degree of freedom can be manipulated upon electrical tuning the valley splitting in Si-based quantum dots[232, 288], encoding qubits based on valley degree of freedom still receives less attention[234].

While in bilayer graphene quantum dots, the controllable valley splitting[173, 174] together with large and electrically tunable valley $g$ factor[177] offer powerful handles for valley manipulations. This enables the realization of spin-valley qubits (as previously demonstrated in carbon nanotube[233]) and valley qubits.

The recent demonstration of spin and valley blockade[191] provides a crucial tool towards qubit manipulation and readout. With such highly tunable and high-quality gate-defined quantum dots, demonstrating a graphene-based qubit is believed to be within reach. It is worth noting that, up to now, it is difficult to tell which type of qubit (spin, valley, or spin-valley) is optimal for realizing quantum dot-based quantum computation applications in bilayer graphene, since those spin-valley states have been mapped very recently and their coherency is still unexplored. As discussed above, further experimental efforts are needed. However, for TMDs systems, theoretical predictions have suggested that it is more promising to use the lowest Kramer's pair to form a spin-valley qubit[113, 119-121], due to the large spin-orbit coupling and relatively small valley $g$ factor.

We also would like to point out that most experiments of 2D materials quantum dots are demonstrated using exfoliated flakes. The weakness in large-scale integration is obvious, compared with conventional materials. Recently, advanced progress has been made in large-area growth of 2D materials, including graphene, hBN, and TMDs (Refs.[292-296]). Gate-controlled quantum dots on these synthesized flakes[209] need to be investigated to address the issue in large-scale device integration.

Besides quantum computation applications, quantum dots are also powerful for mesoscopic physics studies. The diverse family of 2D materials have shown different properties[48, 297, 298], including metallic[45], semiconducting[299], superconducting[300, 301], ferromagnetic[302, 303], and so on. Quantum dots based on them and their heterostructures provide excellent platforms for studying different phenomena at single-particle level, including Kondo effects[189, 190], magneto-Coulomb effects[304-306], and thermoelectric transport[307-310]. A specific example is coupling two quantum dots to a shared superconducting contact to serve as a Cooper-pair splitter[311, 312], which can be regarded as a source of entangled electrons. Furthermore, realizing quantum dots based on topological materials[313-315] and twisted materials[316-318] may also be helpful to unlock the great potentials of the exotic carriers they host.

To conclude, although gate-controlled quantum dots have been demonstrated in a variety of 2D materials, there is still plenty of room in the 2D limit. The 2D material family has great diversity and is easy to hybrid to heterostructures without considering lattice mismatch. These advantages make gate-controlled quantum dots based on 2D materials a promising future, not only for manipulating various quantum degrees of freedom, but also for multi-functional quantum nanodevices.

**Acknowledgements**

This work was supported by the National Natural Science Foundation of China (Grant Nos. 11904351, 61904171, 11625419, 62004185, 12074368, 12034018, 92165207, and 61922074), the Anhui Provincial Natural Science Foundation (Grant Nos. 2008085QF310, and 2108085J03), and the Fundamental Research Funds for the Central Universities (Grant No. WK2030000027).

**Conflict of Interest**

The authors declare no conflict of interest.